\documentstyle[prl,aps]{revtex}           
\begin{document}                          
\preprint{Preprint 5/12/99, submitted to Physics Letters}
\title{Submitted to Physics Letters 5/12/99\\
\   \\
\Large\bf Spin Correlation Coefficients in 
${\boldmath \rm \vec{p}\vec{p}\rightarrow pn\pi^{+}}$
	from 325 to 400 MeV }
\author {Swapan K. Saha,\footnote{permanent address: Bose Institute,
         Calcutta 700009, India} {W.W. Daehnick},\footnote{e-mail :
	 daehnick@vms.cis.pitt.edu} R.W. Flammang,}
\address{Dept. of Physics and Astronomy, University of Pittsburgh,
         Pittsburgh, PA 15260}
\author {J.T. Balewski, H.O. Meyer, R.E. Pollock,
         B. v. Przewoski,
         T. Rinckel, \\
 P. Th\"orngren-Engblom}
\address{Dept. of Physics and Cyclotron Facility, Indiana University,
         Bloomington, IN 47405}
\author {B. Lorentz, F. Rathmann,\footnote{present address:
         Forschungs Zentrum J\"ulich GmbH, 52425 J\"ulich, Germany.}
         B. Schwartz, T. Wise}
\address{University of Wisconsin-Madison, Madison, WI, 53706}
\author {P.V. Pancella}
\address{Western Michigan University, Kalamazoo, MI, 49008}

\maketitle
\vspace{0.5cm}
\begin{abstract}
The spin correlation coefficient 
combinations $\rm S \equiv A_{xx}+A_{yy}$, $\rm D\equiv  A_{xx}-A_{yy}$ and
the 
analyzing powers $\rm A_y(\theta)$ 
were measured for $\rm \vec{p}\vec{p}\rightarrow pn\pi^+$ 
at beam energies of 325, 350, 375 and 400~MeV.  
A polarized internal atomic hydrogen target and a stored, polarized 
proton beam were used. 
These polarization observables are  sensitive to 
contributions of higher partial waves. 
A comparison with recent
theoretical calculations is provided.
\end{abstract}


\twocolumn
\vspace{-0.5cm}

\section{Introduction}
   
   Pion production in nucleon-nucleon collisions has been given
increased attention due to the recent availability of precise
near-threshold data for $pp\rightarrow pp\pi^{0}$, 
$\vec{p}{p}\rightarrow d\pi^{+}$ and $\vec{p}p\rightarrow pn\pi^+$. 
These reactions involve a large momentum transfer, 
but near threshold only a few partial waves can contribute.
Here the study of the short-range features
of the NN interaction is greatly facilitated because it is  possible
to experimentally separate the contributions from different
angular momentum states. In recent studies it was 
found that within 10-20 MeV of threshold,  i.e. below 300 MeV, only Ss 
for  $pp\rightarrow pp\pi^{0}$, \cite{meyer90,meyer92,bonder}  
or at most two partial waves (Ss and Sp) for 
$\vec{p}{p}\rightarrow d\pi^{+}$ and $\vec{p}p\rightarrow pn\pi^+$
\cite{korkmaz,heimberg,drochner,daehnick95,hardie,daehnick98,flammang} 
 were important, and direct comparisons with
simple theoretical approaches can be made. The expected dominance 
of Ss contributions ($l_{NN}=0$, $L_{\pi}=0$) for the final states was
seen, but at the same time it was found that for
$pp\rightarrow pp\pi^0$ traditional models under-predicted the cross 
sections by factors of 3-5. 

At higher energies the reaction $\vec{p}p\rightarrow pn\pi^+$ 
tends to be dominated by the delta resonance.  
\cite{waltham85,shypit89,wicklund87}. 
The present study in the near threshold region will provide information 
on polarization observables in a region still subject to 
relatively parameter-free microscopic calculations. 
Its major motivation is the hope and expectation to reduce the 
theoretical uncertainties for a process fundamental in nuclear physics.
Additional references to prior experimental and theoretical work can be
found in 
\cite{flammang,hardie}.

   Much effort has gone into attempts to explain
 ${pp\rightarrow pp\pi^0}$, probably the theoretically most difficult  
 of the three $\rm NN \rightarrow NN\pi$  branches.
Modern chiral perturbation work on this reaction has been discussed recently
in 
\cite{cohen,sato}, but this approach has not yet explained the data.  
It has become apparent that for $pp \rightarrow pn\pi^+$ 
the first order terms dominate and delta and  complicating heavy 
meson exchange contribution are relatively small making this reaction 
theoretically important.  At this time there is no refereed publication on
chiral 
perturbation calculation; however, very recently a preprint has become 
available that reports $\rm \chi pT$ calculations for $pp \rightarrow
pn\pi^+$
and $pp \rightarrow d\pi^+$.  These $\rm \chi pT $ 
calculations under-predict near threshold data by factors of 2 to 3,
but with less room to adjust poorly known parameters than 
for ${pp\rightarrow pp\pi^0}$ \cite{darocha}. The varying 
theoretical approaches proposed for the inclusion of off-shell
contributions,
``heavy meson''  exchange, and contributions involving the $\Delta$ isobar
are still under discussion \cite{Hanhart98a}.  Divergent results 
suggest that conclusions about the adequacy of a 
 theoretical pion production model cannot be based on agreement for
just a single pion production branch.  A satisfactory model must succeed in 

 explaining all $\rm NN\rightarrow NN\pi$ branches.

Promising attempts to simultaneously describe data from more than one
reaction 
have been made using the Hamiltonian approach . A  relativistic 
phenomenological 
model was offered by Shyam and Mosel \cite{shyam}.  
Their approach naturally included higher angular momenta and
the exchange of heavier mesons and was able to fit $pp\rightarrow pn\pi^+$
and $pp\rightarrow pp\pi^0$ cross sections over a wide energy
range.   This model should also predict analyzing powers; 
however, such predictions have not been published to date.
More complete microscopic calculations in the meson exchange 
picture, involving 
angular momentum states as large as $L_{\pi}=2$ are now being published by
the J\"ulich group \cite{haidenbauer,hanhart97,hanhart98b}.  
Their approach 
of \cite{hanhart97,hanhart98b} is largely  free of adjustable 
parameters.  It gives a very good reproduction of the published 
$pp\rightarrow pp\pi^0$ and $\vec{p} p \rightarrow pn\pi^+$ data.   
However, it  has only qualitative
success in reproducing the new spin correlations observables 
\cite{meyer98} for  
$\vec{p} \vec{p} \rightarrow pp\pi^0$.  The spin correlation data
reported here for $\vec{p}\vec{p}\rightarrow pn\pi^+$  
present a different test of the predictive power
of this approach.

Precision measurements of pion production near threshold became 
possible with the advent of cooled beams and the
technology developed at IUCF and other facilities around the
world to use a stored cooled beam on a windowless internal target. 
In the present experiment we used a polarized beam on a polarized
internal target to measure the $\vec{p}\vec{p}\rightarrow pn\pi^+$
reaction at 325, 350, 375 and 400~MeV.  Bombarding energies for the present
experiment were chosen to cover the range over which some higher 
partial waves gradually become significant. This is the first investigation
of 
spin correlation coefficients in this reaction near threshold.
We have measured $\rm A_{y}$, $\rm S \equiv  A_{xx}+A_{yy}$, 
$\rm D\equiv  A_{xx}-A_{yy}$, 
integrated over all kinematic variables and the observed range of the 
 of the polar angles. We also determined
 the differential analyzing powers $\rm A_y(\theta)$.

\section{Experiment}

The present experiment was conducted at the Cooler ring of the Indiana 
University Cyclotron Facility (IUCF). A beam of polarized protons 
was accelerated to 197~MeV and stack-injected
into the ring.  There it was accumulated for 2-3 minutes,  yielding
orbiting beam currents of 100-300~$\rm \mu$A. The stored beam was then 
accelerated in the ring to full energy. After about 5-10 minutes of data
taking the remaining beam was discarded and the cycle was repeated with
reversed
beam spin. The beam polarization was vertical, alternately up or down.

The target consisted of  $74\pm 4\% $ polarized atomic hydrogen 
in a storage cell, where it was maintained by an atomic beam \cite{wise}.
The open-ended cell,
a movable cylinder of 12~mm diameter with a 25~$\rm \mu$m thick
aluminum wall, minimized the background induced by the beam halo 
in the cell wall. Depolarization
due to wall collisions was minimized by coating the cell with Teflon.
The target gas density distribution along the storage cell was
of triangular shape with a spread of $\pm$12.5 cm about the center.
The thickness of the target was $\rm  \approx 10^{13}~atoms/cm^2$.
The target polarization was flipped every 2 sec, pointing in 
sequence up, down, left, right. Each change of direction took less than
10~ms.

Elastic proton-proton scattering events were observed concurrently with
pion production in order to determine the luminosity and beam 
and target polarization.
Four plastic scintillators at azimuthal angles of 
$\rm \pm45^\circ$ and $\rm \pm135^\circ$  detected 
coincident protons elastically scattered near $\rm \theta_{lab}=45^\circ$.
The 
product PQ of the beam (P) and target (Q) polarization was 
determined from the spin correlation
 $\rm A_{xx}-A_{yy}$, which is large and well known for pp
scattering.  Typically, we saw  $\rm PQ\approx 0.5$, but for the 350 MeV 
run PQ dropped to  0.34.  The time-integrated luminosity was deduced 
from the known pp elastic cross section.
\input{epsf}

\vspace{-4.cm}
\hspace{-3.cm}
\begin{figure}
\epsfysize=10.cm
\centerline{
\epsfbox{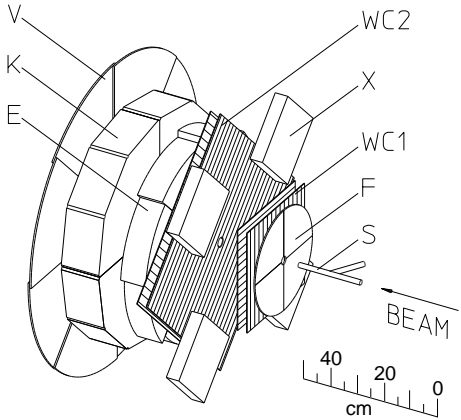}
}
\vspace{-0.0cm}
\caption{Apparatus for the experiment: WC1 and WC2 are wire
chambers. E and K are segmented  plastic scintillators stacks that 
determine the energy of the  charged reaction products. 
V is the veto detector. 
F is a timing detector. X is one of the four  detectors for the elastic pp
scattering
monitor.
\label{fig:Fig1}}
\end{figure}

The detector stack is shown in Fig.~1.
The trajectories of the reaction pions and protons were determined by  wire
chambers WC1 and WC2, spaced 23 cm apart, each consisting of two wire
planes.  
The angular accuracy of the deduced trajectories was 
limited by multiple scattering of the ejectiles and by the 
resolution of the wire chambers. 
The major contribution to the multiple scattering came from the 1.5~mm 
thick plastic scintillator detector (F) which served as timing detector. 
We estimate the angular uncertainty for the  
 trajectories for protons and pions  
of about $\sigma=\rm 0.5^\circ$ and $\sigma=1.0^\circ $, respectively. 
The energy of the charged particles was measured by a stack of segmented  
plastic scintillators of total thickness of 25.4 cm.  
The energy resolution of the stack was about $\rm \sigma=3.5\%$.  
Events are vetoed if a charged particle reaches the detector V behind the   
K-detector. 

Various types of events were analyzed for the present experiment. 
The $\rm \vec{p}\vec{p}\rightarrow pn\pi^{+}$ candidates
had to have simultaneous counts in one or more F segments and two
or more E segments. They were 
rejected if there was a count in the veto detector V. The veto 
detector effectively eliminated (or else selected) very energetic 
(elastic) projectiles.  For 
the 400 MeV run it also rejected the most energetic
reaction protons and pions, complicating the interpretation of this run.
In addition to the pion trigger there were a ``monitor trigger,'' to study
pp
elastic scattering,  and a ``single prong'' trigger for diagnostic purposes.

More details of the experimental set-up are given in \cite{pia}.

\section{Analysis}

Coincident protons and pions were identified by their deposited 
energy and their time of flight
between the E and  F detectors. Software cuts were used to 
further eliminate background. We employed a particle identification 
cut, a cut based on the trajectory origin, and a missing-mass cut.
Fig.~2 shows the distribution of the missing mass reconstructed
from the measured momenta of the proton and pion. 
\begin{figure}
\epsfysize=11.cm
\centerline{
\epsfbox{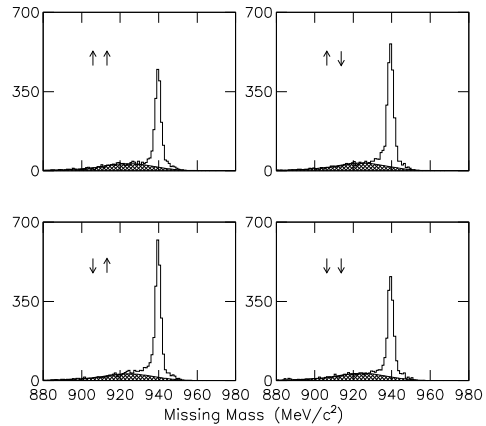}
}
\vspace{-1.5cm}
\caption{Distributions of the calculated mass $\rm m_x$ of the undetected
particle at 325~MeV bombarding energy, shown for the four
combinations of vertical beam and target
polarization. A sharp peak ($\approx \rm 3.5~MeV/c^2$ FWHM) is seen at the
neutron rest mass. The shaded region indicates the
assumed background distribution. }
\end{figure}

The events of interest are those  
near the neutron mass. The  software cuts and the coincidence 
requirements greatly reduce  background, but some of it remains under 
the  neutron mass peak. This amounts to 3 to 8\% of the peak area 
and primarily comes from pions produced in the walls of 
the target cell.  To assess the background shape we took data 
with pure $\rm N_2$ gas in the target cell. We deduced    
the background under the missing mass peak by using a background
shape derived from the $\rm N_2$ missing mass spectrum. This 
empirical shape was normalized to the visible background continuum 
outside the neutron peak. To estimate the error from  background
corrections we varied the background subtraction by $\rm \pm$25\% . 
The effect on the final results was found to be smaller than 
the statistical errors. 
At 325 MeV the resolution of the neutron missing mass peak 
was  $\rm \sigma=1.6~MeV$. 

A Monte Carlo simulation of the experiment was used to determine various
limiting effects of the apparatus and to derive corresponding corrections. 
The code contained the detailed geometry
of the detector systems and the target density distribution. It included the
effects
of multiple scattering of the charged particles, pion decay, loss of energy
by the charged particles, and the energy resolutions of the scintillators. 
The simulation provided a guideline for the expected angular and energy 
distributions of pions and protons. This was important at 400 MeV, 
because about 0.2\% of the coincident protons and 14.4\% of the pions, 
intercepted by the detectors, did not stop 
in the scintillators. In the final analysis we used only those 
events where the proton and the pion energies were measured accurately 
enough to produce a  missing mass value within $\rm \pm 3.5~MeV$ of the 
neutron mass. The simulation was also needed 
to study the acceptance of the detector system, which  did
not cover the entire phase space of the exit channel.
Very forward events were suppressed by the central hole in the detector 
which was needed to accommodate the beam pipe for the circulating 
beam.  Many large-angle events are lost because of the finite detector size.
Coincident events are detected only if the polar angle of the 
proton is larger than $\rm 6^\circ$
and if the pion angle is $\rm 6^\circ \leq\theta_{lab}\leq 32^\circ$. 
The event acceptance for $\rm p\pi^{+}$ coincidences 
was 21.5\% at 325 MeV,  19.4\%  at 350 MeV, 18.2\%  at 375 MeV, and 15.0\%  
at 400 MeV.  The resulting
loss in count rate is significant;
however, since target and projectile are identical particles
we sample a  phase space twice as large as the acceptance 
suggests.
 
In the c.m. system low energy pions are seen for all 
angles, but energetic pions are not sampled for polar angles near 
$\theta_{c.m.}=90^{\circ}$.  
Near threshold the analyzing power for pions in the $\rm p n \pi^{+}$ 
final state  varies about linearly with pion momentum, independent of 
bombarding energy \cite{daehnick98}.
Therefore $\rm A_{y}$ data for c.m. angles between 
$\rm65^{\circ} \ and \ 115^{\circ}$ not 
only have low statistics, but they also require (model dependent) 
corrections, which results in large error bars.
The integrated spin dependent cross section ratios S, D, and $A_{y}$ 
reported here lack data near $\theta_{c.m.}=90^{\circ}$, 
and therefore only approximately represent the full integral over $\rm 
\theta_{c.m.}$. 
Nevertheless, accurate comparisons with theory are possible.  They 
require a  corresponding integration range for the theoretical calculation,
i.e., the 
restriction of pion lab angles to the range  $\rm 6^{\circ}$ to
$32^{\circ}$.

\section{Results and Discussion}

Fig.~2 compares four missing mass spectra with the same integrated 
luminosity for the four spin-spin orientations with vertical beam and 
target polarization.
 The spin-dependent  cross section $\Delta\sigma'_T$ 
 (defined as the observed cross sections with 
the spins of the colliding protons opposite minus the 
observed cross section with spins parallel) can be calculated 
by using the cross ratio method\cite{meyer97b}: 
\begin{equation}
\rm \frac{\Delta\sigma'_T}{\sigma_{unpol}} = \frac{2}{PQ} \
\frac{1-\sqrt{R}}{1+\sqrt{R}}
\end{equation}
where $\rm R = (Y_{uu}Y_{dd}/Y_{ud}Y_{du})$.  $\rm Y_{mn}$ 
denotes the yields 
for the four spin combinations  and  $\rm \sigma_{unpol}$ is the 
unpolarized cross section for the measured  range of the polar angle 
$\rm \theta$. 
Because of the limited $\rm \theta$ range of the detector the integrated
cross 
section ratio $\rm \Delta\sigma'_T/\sigma_{unpol}$ will differ from
the standard ratio $\rm \Delta\sigma_T/\sigma_{total}$ if the spin 
correlation coefficients have a strong $\rm \theta$  dependence. 
The recent Hamiltonian calculation by Hanhart et al. \cite{hanhart98b}
suggests relatively small changes for our geometry (see Fig.~4).

For the present experiment spin-dependent differential cross section  
as a function of the pion azimuthal angle $\rm \phi_{\pi}$
can be expressed as\cite{meyer97b}
\begin{eqnarray}
\sigma(\rm \phi_{\pi}) = \sigma_{tot}[1+(p_yA_y^B - q_yA_y^T)\cos\phi_{\pi}
\nonumber \\
  +p_yq_y(\frac{A_{xx}+A_{yy}}{2}-\frac{A_{xx}-A_{yy}}{2}\cos2\phi_{\pi})
\nonumber\\
  + terms\ involving\  p_x,p_z,q_x,q_z],
\end{eqnarray}
where $\rm A_y^B$ and $\rm A_y^T$ are respectively the beam and target 
analyzing powers, and $\rm p_x,p_z,q_x,q_z$ are very small.
The values $A_{nn}$ are the spin correlation coefficients and 
$\rm p_i(q_i)$ are the beam (target) polarization components. In this
equation
the observables $\rm A_y^B$, $\rm A_y^T$ and $A_{nn}$ are integrals
over all kinematic variables of the proton, the neutron and the pion 
except over the pion azimuthal angle $\phi_{\pi}$. Spin correlation
observables 
are deduced by sorting the
data as a function  $\rm \phi_{\pi}$ for a specific
combination of beam and target spin. 
We show the $\phi$-dependence of the measured ratios in Fig. 3.

Except for the scales, the presentation 
is identical to that in ref.\cite{meyer98} for $\rm 
\vec{p}\vec{p}\rightarrow  pp\pi^{0}$.
The columns represent the four bombarding energies,
and the rows the different combinations $\rm R_i$ for the spin-dependent
yields. The definitions of the $\rm R_i$ are given in Fig. 3.
To obtain the plotted quantities, we calculate
$\rm W_i = (R_i-1)/(R_i+1)$ and divide by $\rm \sqrt{PQ}$ for the first 
three rows and by PQ for the last two two rows.

\onecolumn 
\begin{figure}
\epsfysize=20.0cm
\centerline{
\epsfbox{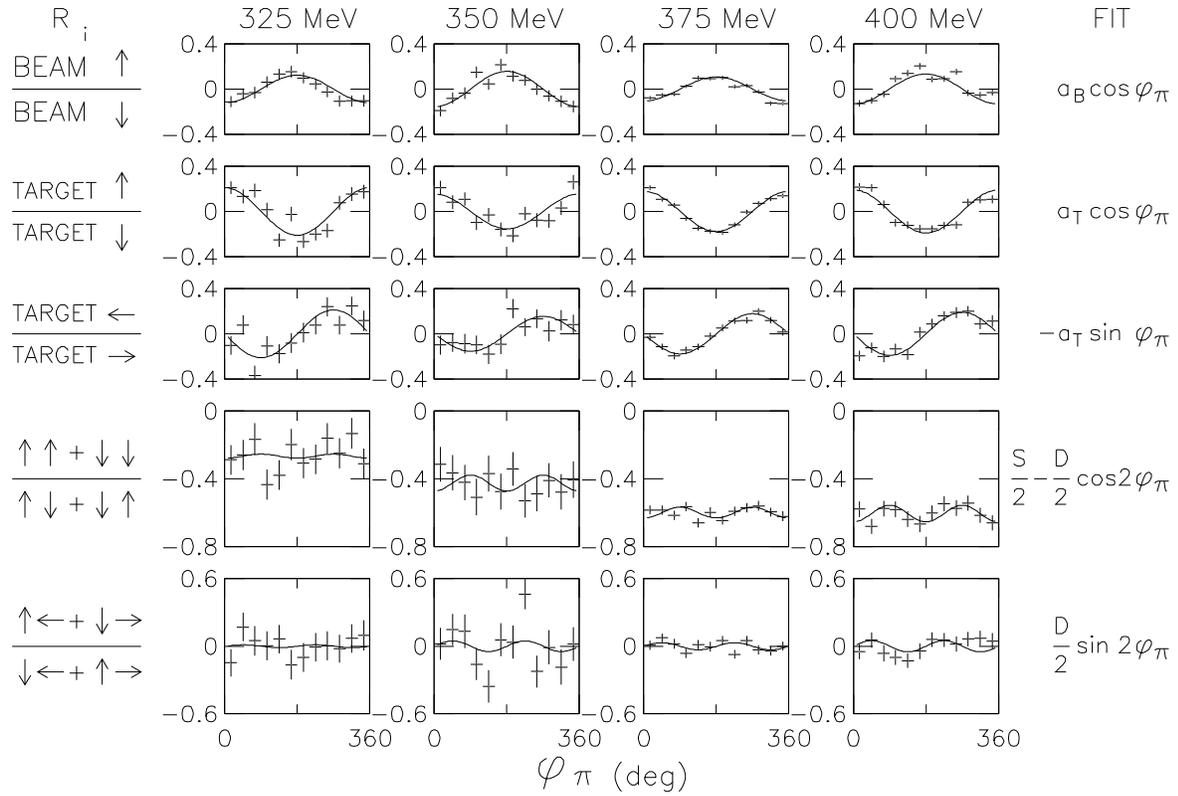}
}
\vspace{-4.0cm}
\caption{Asymmetries for different spin combinations $\rm R_i$ (listed on
the
left) as a function of the azimuthal angle of the $\rm \pi^+$. The solid
curves represent a least-square fit using the theoretical
$\rm \phi_{\pi}$ dependence (listed on the right), varying $\rm a_B$,
$\rm a_T$, $\rm S\equiv (A_{xx}+A_{yy})$, and $\rm D\equiv
(A_{xx}-A_{yy})$.}
\end{figure}
\twocolumn 

The first three rows are directly related to the beam and target
analyzing powers, whereas the 
last two rows reflect the spin correlation coefficient combinations 
$\rm S\equiv (A_{xx}+A_{yy})$, and $\rm D\equiv (A_{xx}-A_{yy})$. In this 
figure $\rm a_B$ and $\rm a_T$ are related to the beam and target analyzing 
powers $\rm A_y^B$ and $\rm A_y^T$ by $\rm a_B=A_y^BP/\sqrt{PQ}$, and 
$\rm a_T=A_y^TQ/\sqrt{PQ}$.  
The solid lines in Fig.~3 represent a least-square fit using the  
$\rm \phi_{\pi}$ dependence from Eq.(2) (listed to the right of Fig.~3), 
and varying the four parameters $\rm a_B$, $\rm a_T$, S, and D. 

\begin{figure}
\epsfxsize=5.5cm
\centerline{
\epsfbox{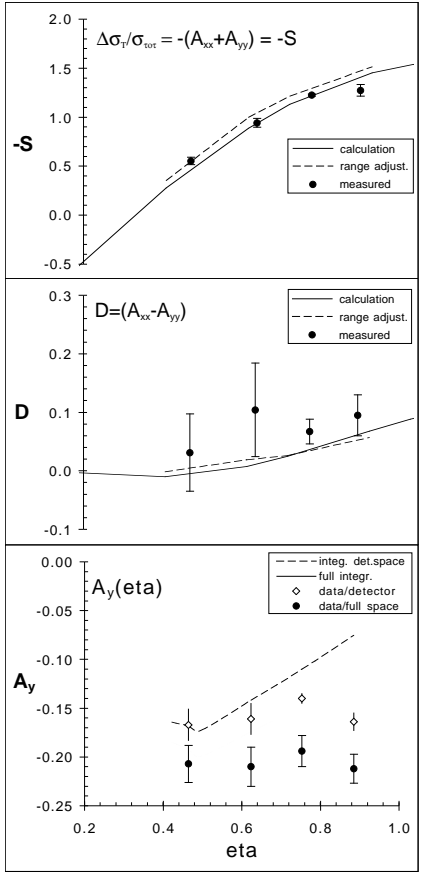}
}
\vspace{+0.5cm}
\caption{The integrated $\rm \vec{p}\vec{p}\rightarrow pn\pi^+$
 observables
$\rm \Delta\sigma'_T/\sigma_{unpol} = -(A_{xx}+A_{yy}$),
$\rm D\equiv A_{xx}-A_{yy}$, and $\rm A_y$ as listed in Table~1 are shown
as a function of  $\rm \eta$, the maximum
pion center-of-mass momentum in units of the pion mass. The
observables $\rm A_{xx}$, $\rm A_{yy}$ and $\rm A_{y}$ were
integrated over all kinematic variables except the
pion azimuthal angle. Solid curves are from Hanhart et al.. 
These calculations are for the full phase space and are not directly
comparable
with the data. The dashed curves approximate the theoretical calculations
for the actual phase space observed.} 
\end{figure}

The integrated spin correlation coefficient S is related to the 
spin-dependent  cross section by  $\rm \Delta\sigma'_T/\sigma_{tot} = -S$, 
and we find that these two analysis methods agree. 
The integrated analyzing power is 
given by $\rm A_y=(-a_B\cdot a_T)^{1/2}$. The resulting polarization 
observables are shown in Fig.~4 and listed in Table~1. These values, 
but not those in Fig.~3, are corrected for background contributions. 
The values for the integrated $\vec{p}\vec{p}\rightarrow pn\pi^+$ 
spin correlation coefficients in Fig.~4  are in remarkably 
good agreement with recent 
Hamiltonian predictions following ref. \cite{hanhart98b}.\\

In our experiment beam and target analyzing powers 
are related by  $\rm A_y^B(\theta)=-A_y^T(\pi-\theta) $ and,
therefore, permit a study of the back angles for  $\rm A{_y}(\theta_{c.m.})$
with the present apparatus.
Fig.~5 shows the observed $\theta_{c.m.}$ dependence of the analyzing 
powers $A_{y}(\theta_{c.m.})$.  As mentioned above, 
the limited size of the detector reduced the usefulness of data
 near $\theta_{c.m.}= 90^{\circ}$.  However, a large enough 
range in $\theta$ could be covered to demonstrate the change in the 
angular distributions with energy and the need for  
partial waves higher than Ss and Sp.
The dashed curves show calculations taken from \cite{hanhart98b}.
\begin{figure}
\epsfxsize=5.5cm
\centerline{
\epsfbox{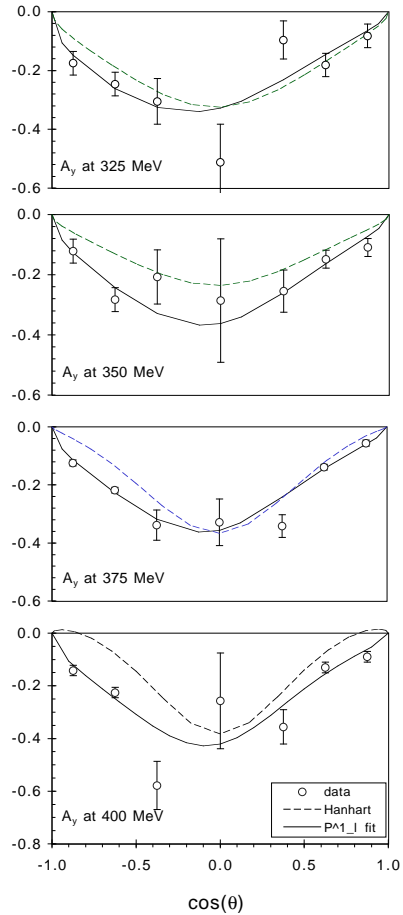}
}
\vspace{0.1cm}
\caption{The measured $\rm \vec{p}\vec{p}\rightarrow pn\pi^+$
analyzing powers $\rm {A_{y}}(\theta_{c.m.}, T)$ compared with fits (solid
curves) described in the text. The angular distributions suggest some
asymmetry 
around $\rm \theta_{c.m.}=90^{\circ}$. Acceptable fits require the 
inclusion of an associated Legendre polynomial with $L_{\pi}=2$. The 
detector size did not permit reliable measurements for 
$\rm -0.25 < cos(\theta_{c.m.}) < 0.25$. The dashed curves are calculations
by Hanhart et al.[18].}

\end{figure}

The solid curves in Fig. 5 represent fits with the relation
\begin{equation}
A_{y}(x)=(1/\sigma_{unpol}(x))*(a~P^1_{1}(x)
+b~P^1_{2}(x)+c~P^1_{3}(x)), 
\end{equation}
where the expressions ${P^1}_{\nu}(x)$ are 
associated Legendre polynomials and $x=cos(\theta)$.  
$\rm P^1_{3}(x)$ was not needed or used for
the fits shown; however,  $\rm P^{1}_{2}(x)$,
i.e. the inclusion of an $L_{\pi}=2$ term, was essential to obtain a 
reasonable representation of the data at the four energies.

We have found that the recent calculations of the J\"ulich group
are quite successful in predicting the integrated spin correlations 
for $\vec{p}\vec{p}\rightarrow pn\pi^+$ in spite of their difficulties
with $\vec{p}\vec{p}\rightarrow pp\pi^\circ$.  This dramatic difference 
in the otherwise similar reactions should help identify the reason for the
less 
than satisfactory agreement of calculations and measurements
\cite{meyer98} for $\vec{p}\vec{p}\rightarrow pp\pi^\circ$.
A further indication of remaining shortcomings is found in the
difference between calculated and measured angular distributions 
for  $\rm A_{y}(\theta_{c.m.})$ in Fig. 5. It appears that some 
diagrams that are secondary in $\vec{p}\vec{p}\rightarrow pn\pi^+$ 
but important in $\vec{p}\vec{p}\rightarrow pp\pi^\circ$ have been 
omitted or treated inadequately.

We are currently analyzing additional measurements of the 
$\rm \vec{p}\vec{p}\rightarrow pn\pi^+$ reaction with
longitudinal beam and target polarization. Such  results
together with the present data may allow a deduction of
the dominant partial waves in a model independent way.\\


\begin{acknowledgements}
    We acknowledge the support of Drs.~M.~Dzemidzic, F.~Sperisen 
and D.~Tedeschi in the early stages of the experiment.
We thank Dr. W. Haeberli for his advice and continuing interest and J.
Doskow for technical support.  We also
wish to thank the IUCF accelerator operations group for their
dedicated efforts. We are grateful to the authors of
ref. \cite{hanhart98b} for  making available to us calculations
for $\rm \vec{p}\vec{p}\rightarrow pn\pi^+$
obtained with their model.
This work was supported by the US National Science Foundation under
Grants PHY95-14566, PHY96-02872, PHY-97-22556, and by the department
of Energy under Grant DOE-FG02-88ER40438.
\end{acknowledgements}

\pagebreak

\onecolumn 
\begin{table}
\caption{Product of beam and target polarization and the 
deduced integrated spin correlation coefficients. The observables 
listed are affected differently by the limited 
observed range for the polar angle $\rm \theta$.  Model calculations show
that 
the lack of counts for the range 
$\rm -0.4 < cos(\theta_{c.m.}) < 0.4$ makes the listed values for -S and D 
larger than what a complete integral would have produced.
The effect is more pronounced for $\rm A_{y}$ 
because $\rm A_{y}(\theta_{c.m.})$ has a strong minimum near $\rm 
\theta_{c.m.} =90^{\circ}$.  The last column provides the full integral over
the fits to 
$A_{y}(\theta)$ in Fig. 5 and a better value for $\rm A_{y}$.  See text.}
\label{table1}
\begin{tabular}{ccccccc}
T &   $\rm \eta$  &  $\rm P\cdot Q$  &$-S \equiv$   &
$\rm D \equiv $  &  $\rm A_y$  &$\rm A_y $\\
 (MeV) & & &  $\rm -(A_{xx}+A_{yy})$ &$\rm A_{xx}-A_{yy}$ &(detector space)
& 
 (full space, Fig.5)  \\
 \tableline

325.6&0.464& 0.456$\pm$0.003& 0.553$\pm$0.038&  
0.031$\pm$0.066&$-$0.167$\pm$0.016&-0.207$\pm$ 0.019\\
350.5&0.623& 0.342$\pm$0.004& 0.942$\pm$0.046&  
0.104$\pm$0.080&$-$0.161$\pm$0.016&-0.210$\pm$ 0.020\\
375.0&0.753& 0.514$\pm$0.004& 1.226$\pm$0.011&  
0.067$\pm$0.021&$-$0.140$\pm$0.005&-0.194$\pm$ 0.016\\
400.0&0.871& 0.526$\pm$0.006& 1.274$\pm$0.019&  
0.095$\pm$0.035&$-$0.164$\pm$0.009&-0.212$\pm $0.015
\end{tabular}
\end{table}

\end{document}